\documentclass[conference]{IEEEtran}
\usepackage{cite}

\ifCLASSINFOpdf
\else
  \usepackage[dvips]{graphicx}
\fi
\usepackage[cmex10]{amsmath}
\usepackage{array}
\usepackage[tight,footnotesize]{subfigure}
\usepackage{url}
\usepackage{amsthm}
\usepackage{amssymb}
\theoremstyle{plain}
\newtheorem{lem}{\textbf{Lemma}}

\newtheorem{corollary}{Corollary}
\theoremstyle{remark}

\begin{document}
%
\title{On the Role of Feedback in LT Codes}

\author{
\IEEEauthorblockN{
Jesper H. S\o rensen\IEEEauthorrefmark{1},
Petar Popovski\IEEEauthorrefmark{1},
Jan Ostergaard\IEEEauthorrefmark{1},}
\IEEEauthorblockA{\IEEEauthorrefmark{1}Aalborg University, Department of Electronic Systems, E-mail: \{jhs, petarp, jo\}@es.aau.dk}
}


\maketitle

\begin{abstract}
This paper concerns application of feedback in LT codes. The considered type of feedback is acknowledgments, where information on which symbols have been decoded is given to the transmitter. We identify an important adaptive mechanism in standard LT codes, which is crucial to their ability to perform well under any channel conditions. We show how precipitate application of acknowledgments can interfere with this adaptive mechanism and lead to significant performance degradation. Moreover, our analysis reveals that even sensible use of acknowledgments has very low potential in standard LT codes. Motivated by this, we analyze the impact of acknowledgments on multi layer LT codes, i.e. LT codes with unequal error protection. In this case, feedback proves advantageous. We show that by using only a single feedback message, it is possible to achieve a noticeable performance improvement compared to standard LT codes.
\end{abstract}
\IEEEpeerreviewmaketitle

\section{Introduction}
Rateless codes are capacity achieving erasure correcting codes. Common for all rateless codes is the ability to generate a potentially infinite amount of encoded symbols from $k$ input symbols. Decoding is possible when $(1+\epsilon)k$ encoded symbols have been received, where $\epsilon$ is close to zero. The generation of encoded symbols can be done on the fly during transmission, which means the rate of the code decreases as the transmission proceeds, as opposed to fixed rate codes, hence the name. Rateless codes are attractive due to their flexible nature. Regardless of the channel conditions, a rateless code will approach the channel capacity, without the need for feedback, except a notification when decoding is successfully completed. Moreover, practical implementations of rateless codes can be made with very low encoder and decoder complexity. The most successful examples are LT codes \cite{fc2} and Raptor codes \cite{raptor}. LT codes are the focus of this work.

Although LT codes are independent of feedback, a feedback channel might still be available in the network. This is a resource which can be exploited to potentially improve LT coded transmissions. The question of how to exploit such a feedback channel has recently received attention. An approach called Real-Time oblivious erasure correcting is proposed in \cite{feedfc1}. This approach utilizes feedback telling the transmitter how many of the $k$ input symbols have been decoded. With this information the transmitter chooses a fixed degree (explained in section \ref{background}) for future encoded symbols, which maximizes the probability of decoding new input symbols. Shifted LT codes \cite{feedfc3} is another application of this type of feedback. Here the feedback is used to shift the degree distribution towards higher degrees as decoding proceeds, which they claim is desirable in LT codes. In \cite{feedfc2}, another approach called doped fountain coding is proposed. In this approach the receiver feeds back information on the current buffer content, i.e. undecoded encoded symbols. This makes the transmitter able to select and transmit the input symbols, which will accelerate the decoding process the most.

In this work, feedback in the form of acknowledgments, ACKs, is considered. We will analyze its impact on LT codes, both for prioritized and non-prioritized data. In the case of prioritized data, we analyze rateless codes with unequal error protection \cite{uepfc1} in combination with our proposed application of feedback. We argue that this combination holds very desirable synergies. Another important contribution of this work is the identification and description of an inherent adaptive mechanism in LT codes. That mechanism makes LT codes able to adapt to the current decoder state, without the need for feedback. This is a remarkable property, and it is the sole reason why LT codes succeed in approaching the channel capacity, regardless of the channel conditions. When applying feedback, it is important to bear the inherent adaptive mechanism in mind, because any actions based on the feedback should not negatively interfere with it. This will be evident from our evaluations.

The remainder of this paper is structured as follows. Section \ref{background} gives an introduction to LT codes, explaining the encoding and decoding processes and the relevant terms. The analytical work is described in section \ref{sec:analysis}, followed by a numerical evaluation of the investigated schemes in section \ref{sec:results}. Conclusions are drawn in section \ref{sec:conclusion} before an Appendix finalizes the paper.


\section{Background} \label{background}
In this section an overview of standard LT codes is given. Assume we wish to transmit a given amount of data, e.g. a file or slice of video from a stream. This data is divided into $k$ \textit{input symbols}. From these input symbols a potentially infinite amount of encoded symbols, also called \textit{output symbols}, are generated. Output symbols are XOR combinations of input symbols. The number of input symbols used in the XOR is referred to as the \textit{degree} of the output symbol, and all input symbols contained in an output symbol are called \textit{neighbors} of the output symbol. The output symbols of an encoder follow a certain degree distribution, $\pi(i)$, which is a key element in the design of good LT codes. The encoding process of an LT code can be broken down into three steps:

\textbf{Encoder:}
\begin{enumerate}
\item Randomly choose a degree $i$ by sampling $\pi(i)$.
\item Choose uniformly at random $i$ of the $k$ possible input symbols.
\item Perform bitwise XOR of the $i$ chosen input symbols. The resulting symbol is the output symbol.
\end{enumerate}

\noindent This process can be iterated as many times as needed, which results in a rateless code.

A widely used decoder for LT codes is the belief propagation (BP) decoder. The strength of this decoder is its very low complexity \cite{raptor}. It is based on performing the reverse XOR operations from the encoding process. Initially all degree one output symbols are identified and moved to a storage referred to as the \textit{ripple}. Symbols in the ripple are \textit{processed} one by one, which means that they are removed as content from all buffered symbols through XOR operations. Once a symbol has been processed, it is removed from the ripple and considered decoded. The processing of symbols in the ripple will potentially reduce some of the buffered symbols to degree one, in which case they are moved to the ripple. This is called a symbol \textit{release}. This makes it possible for the decoder to process symbols continuously in an iterative fashion. The iterative decoding process can be explained in two steps:

\textbf{Decoder:}
\begin{enumerate}
\item Identify all degree one output symbols and add them to the ripple.
\item Process a symbol from the ripple and remove it afterwards. Go to step 1.
\end{enumerate}

Decoding is successful when all input symbols have been recovered. If at any point before this, the ripple size equals zero, decoding has failed. If this happens, the receiver can either signal a decode failure to the transmitter, or wait for more output symbols. In the latter case, new incoming output symbols are initially stripped for already recovered input symbols at the receiver. If the resulting output symbol has degree one, the iterative decoding process is restarted.

\subsection{Acknowledgments in LT Codes}
The type of feedback considered in this paper is ACKs. In LT codes this can be used to inform the transmitter, which input symbols have been decoded at the receiver. We propose to act on this by excluding the ACK'ed symbols from future encoding. More specifically this has the following impact on the encoding process: In step 1 of the encoder, the new decreased number of input symbols is taken into account in the degree distribution. Thus the parameter $k$ is updated to $k'=k-M$, where $M$ is the number of ACK'ed symbols. In step 2, the $d$ input symbols are chosen among the $k'$ undecoded input symbols. ACKs have no impact on the BP decoder. The cost of transmitting the ACKs is not taken into account in this work.


\section{Analysis} \label{sec:analysis}
The purpose of this analysis is to obtain insight on the impact of ACKs in LT codes. The performance metric considered in this work is the average overhead. Thus, especially the impact on the probability of receiving redundant symbols is interesting. Initially single layer LT codes, i.e. standard LT codes for non-prioritized data, are analyzed. The results from this motivate us to extend the analysis to multi layer LT codes for prioritized data, which is done in section \ref{sec:multilayer}.

\subsection{ACK in Single Layer LT Codes}
The output of the encoder is a symbol of degree $i$. However, such a symbol might include already decoded data, which is removed upon arrival through XOR operations. As a result, the degree of the received symbol is reduced, before it is included in the iterative BP decoding process. We denote the \textit{reduced degree} $i'$ and its distribution $\pi'(i')$.

\begin{lem}(Reduced Degree Distribution)\label{rdd}
Given that the encoder applies $\pi(i)$, and that $L$ out of $k$ input symbols remain undecoded, the reduced degree distribution is found as

\begin{align}
\pi'(i') &= \sum_{i=i'}^{i'+k-L} \left( \pi(i) \cdot \frac{\binom{L}{i'}\binom{k-L}{i-i'}}{\binom{k}{i}} \right) &\hspace{0.2cm} \mathrm{for} \hspace{0.2cm} 0 \le i' \le L, \notag \\
\pi'(i') &= 0 &\hspace{0.2cm} \mathrm{for} \hspace{0.2cm} L < i' \le k. \notag
\end{align}
\end{lem}

\begin{IEEEproof}
A degree reduction from $i$ to $i'$ happens when $i'$ of the neighbors are among the $L$ undecoded symbols and the remaining $i-i'$ neighbors are among the $k-L$ already decoded symbols. The probability of this event can be evaluated using the hypergeometric distribution. The probability that a symbol of reduced degree $i'$ is received is the sum of all contributions from symbols of original degree $i=\{i',i'+1,...,i'+k-L\}$.
\end{IEEEproof}

Lemma \ref{rdd} basically says that the probability of receiving a symbol with reduced degree $i'$ is a weighted sum of the part of the original degree distribution where $i'\le i \le i'+k-L$. This tells us that the applied degree distribution experiences a shift towards lower degrees when a subset of the $k$ input symbols has been decoded. This shift grows stronger as the number of decoded symbols increases. Thus, although the original degree distribution is constant during transmission, $\pi'(i')$ is dynamic, since it depends on $L$. The increasing shift of the reduced degree distribution is an important element of LT codes. The reason is that low degree symbols are preferable late in the transmission. New arriving symbols can serve one of two purposes. Either $i'\geq2$ and the symbol is stored in the buffer, or $i'=1$ and the iterative decoding process is initiated with the new known input symbol. In short, new symbols either build up potential or release potential. Late in the transmission, when much potential has been built up, it is preferable to start releasing this potential instead of building up even more. This is achieved by the inherent shift of $\pi'(i')$, which lets the LT code adapt to the current decoder state, without the need for feedback.

\begin{figure}[t]
\centering
\includegraphics[width=0.95\columnwidth]{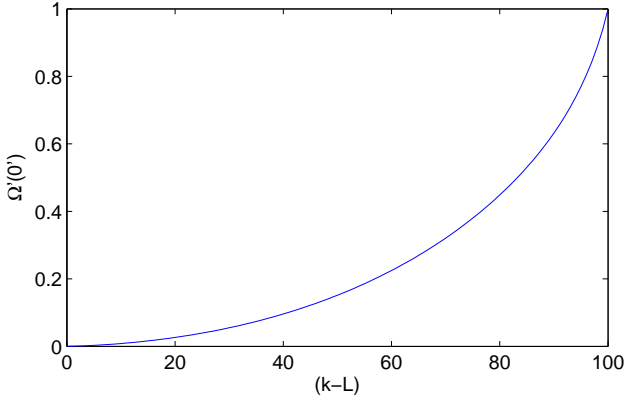}
\caption{The probability of receiving a symbol with reduced degree zero, as a function of $(k-L)$, when applying the RSD at $k=100$.}
\label{fig:onelayerred}
\end{figure}

In the case that the degree is reduced to zero, the symbol is redundant. Thus, $\pi'(0)$ is an important parameter when trying to minimize the overhead of the LT code. It has been plotted as a function of $k-L$, i.e. the number of decoded symbols, in Fig. \ref{fig:onelayerred}. The applied degree distribution is the Robust Soliton distribution, RSD, and $k=100$. The RSD is the de facto standard degree distribution for LT codes and was originally proposed in \cite{fc2}. The figure reveals that as the decoding progresses, the probability that a symbol will be redundant increases exponentially. We would like to explore how this probability can be decreased, by applying feedback in the form of ACK.

\begin{lem}(Acknowledgments Reduce Redundancy)
$\pi'(0)$ is a strictly decreasing function of $M$, $0\le M \le k-L$, where $M$ denotes the number of ACK'ed input symbols. 
\end{lem}

\begin{IEEEproof}
When $i'=0$ and the decrease of $k$, caused by the $M$ ACK'ed symbols, is taken into account, the equation in Lemma \ref{rdd} reduces to

\begin{align}
\pi'(0) &= \sum_{i=0}^{k-M-L} \left( \pi(i) \cdot \frac{\binom{k-M-L}{i}}{\binom{k-M}{i}} \right). \notag
\end{align}

\noindent The fraction of binomial coefficients is a strictly decreasing function of $M$, because

\begin{align}
\scriptstyle\binom{k-M-L}{i}-\binom{k-(M+1)-L}{i} &\scriptstyle < \binom{k-M}{i}-\binom{k-(M+1)}{i}, \notag \\
\scriptstyle\frac{(k-M-L)!}{i!(k-M-L-i)!}-\frac{(k-M-1-L)!}{i!(k-M-1-L-i)!} &\scriptstyle < \frac{(k-M)!}{i!(k-M-i)!}-\frac{(k-M-1)!}{i!(k-M-1-i)!}, \notag \\
\scriptstyle\prod\limits_{j=0}^{i-1}(k-M-L-j)-\prod\limits_{j=0}^{i-1}(k-M-1-L-j) &\scriptstyle < \prod\limits_{j=0}^{i-1}(k-M-j)-\prod\limits_{j=0}^{i-1}(k-M-1-j), \notag \\
\scriptstyle i\prod\limits_{j=1}^{i-1}(k-M-L-j) &\scriptstyle < i\prod\limits_{j=1}^{i-1}(k-M-j),\hspace{0.1cm} \forall M, \hspace{0.05cm} 0\le M \le k-L, \notag
\end{align}
\noindent since L is positive, and from this it follows that $\pi'(0)$ is a strictly decreasing function of $M$.
\end{IEEEproof}

Unfortunately, reducing the probability of redundancy is not the only consequence of ACK'ing symbols. It also has an impact on the rest of the reduced degree distribution. In general the reduced degree distribution will experience a shift towards higher degrees, when symbols are ACK'ed. In this way, ACK'ing data acts as a counter to the desirable dynamic property of the reduced degree distribution. In fact, when the maximum number of ACK's is made, the dynamic property is eliminated entirely.

\begin{lem}(Acknowledgments Counter Dynamic Reduced Degree)\label{counter}
When $M=k-L$, the reduced degree distribution equals the original degree distribution.
\end{lem}

\begin{IEEEproof}
\begin{align}
\pi'(i') &= \sum_{i=i'}^{i'+k-M-L} \left( \pi(i) \cdot \frac{\binom{L}{i'}\binom{k-M-L}{i-i'}}{\binom{k-M}{i}} \right) \notag \\
         &= \sum_{i=i'}^{i'} \left( \pi(i) \cdot \frac{\binom{L}{i'}\binom{0}{i-i'}}{\binom{L}{i}} \right) \notag \\
         &= \pi(i') \notag
\end{align}
\end{IEEEproof}

Thus, we conclude that the proposed application of feedback in the form of ACK's has both a positive and a negative effect on a standard LT code. On one hand, it reduces the probability of receiving redundant symbols. On the other hand, it eliminates an important inherent adaptive mechanism of LT codes. We propose to solve this problem by applying an adaptive original degree distribution. From Lemma \ref{counter} we saw that when all decoded symbols have been ACK'ed, the reduced degree distribution follows the original degree distribution. Hence, by implementing a dependency on $L$ in the original degree distribution, we can restore the desired adaptive mechanism. More specifically, we should apply the distribution from Lemma \ref{rdd}, with the small adjustment that $\pi(0)=0$, because we obviously do not want to generate symbols of degree zero. Moreover, since we truncate the distribution, we must normalize. The above observations can be summarized in the following corollary.

\begin{corollary}(Feedback Based Adaptive Degree Distribution)
Under the assumption that all decoded symbols have been ACK'ed by the receiver, the following defines a distribution which avoids symbols of reduced degree zero and preserves the adaptive mechanism of LT codes:
\begin{align}
\rho(i) &= \sum_{j=i}^{i+k-L} \frac{\pi(j) \cdot \binom{L}{i}\binom{k-L}{j-i}}{(1-\pi'(0))\binom{k}{j}}, &\hspace{0.2cm} \mathrm{for} \hspace{0.2cm} 1 \le i \le L. \notag
\end{align}
\begin{flushright}
$\blacktriangledown$
\end{flushright}
\end{corollary}

A simulation has been performed with the purpose of evaluating the application of ACKs in standard LT codes. Three different schemes are compared; A standard LT code without feedback using the RSD, an LT code with feedback using the RSD and an LT code with feedback using $\rho(i)$, where $\pi(j)$ is the RSD. For the schemes utilizing feedback, it is assumed that each individual input symbol is ACK'ed immediately after it has been decoded and that the ACK is received at the server before the next transmission. In this way, the server will know which symbols have been decoded at all encoding times. The cost of transmitting the feedback is not taken into account. These are ideal conditions, thus the results are upper bounds for these applications of feedback. The parameters of the RSD are $c=0.1$ and $\delta=1$, since these have been found to provide the smallest average overhead in \cite{lc5}. The simulations have been performed at $k=1000$ and the results are averages of 1000 simulations.

\begin{figure}[t]
\centering
\includegraphics[width=0.95\columnwidth]{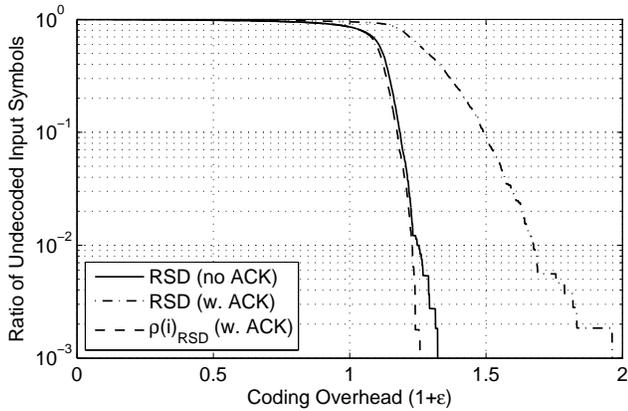}
\caption{The relative number of undecoded input symbols as a function of the amount of received symbols for the three simulated schemes.}
\label{fig:noackvsidealack}
\end{figure}

Fig. \ref{fig:noackvsidealack} shows the results where the relative number of undecoded input symbols has been plotted as a function of the amount of received symbols. From the figure it is seen that when applying ACKs without changing the original degree distribution, the performance degrades significantly. Thus, the negative effect from ACK's by far outweighs the positive effect. However, when we eliminate the negative effect by implementing an adaptive original degree distribution, we see a performance increase compared to the standard LT code with no feedback. 

The gain from introducing feedback in the form of ACKs is quite small though. This is explained by the very rapid change in LT codes from not having decoded anything to having decoded everything. As mentioned earlier, an LT code initially builds up potential and then eventually releases this potential. As a result, very few input symbols are decoded in the beginning and then suddenly, within relatively few new received encoded symbols, all input symbols are decoded. This is sometimes referred to as \textit{avalanche decoding} \cite{mackay}. This leaves a very narrow window in which ACKs are applicable, hence the small performance gain.

For this reason it is particularly interesting to investigate the use of ACKs in combination with multi layer LT codes, also referred to as LT codes with Unequal Error Protection, UEP \cite{uepfc1}. In these codes the aim is to successfully decode the data in stages, i.e. layers are decoded at different times or equivalently with different average overheads. This leaves a longer window for the use of ACKs, which potentially increases the gain.

\subsection{ACK in Multi Layer LT Codes} \label{sec:multilayer}
For the investigation of the use of ACK in multi layer LT codes we will use the approach to layering proposed in \cite{uepfc1}. This is a simple approach, where the uniform distribution used for selection of input symbols is replaced by a distribution which favors symbols from more important layers. In general, the $k$ input symbols are divided into $N$ layers $s_1$, $s_2$,..., $s_N$, each with size $\alpha_1 k$, $\alpha_2 k$,..., $\alpha_N k$, where $\sum_{j=1}^N \alpha_j = 1$. The probability of selection associated with input symbols from $s_j$ is denoted $p_j(k)$, such that $\sum_{j=1}^N p_j(k)\alpha_j k = 1$. Note that if $p_j(k)=\frac{1}{k}$ $\forall j$, then all data is treated equally, as in the single layer LT code. In this analysis, only two layers are considered, which are referred to as the base layer and the refinement layer, respectively. We parameterize the favoring of the base layer as $\beta=\frac{p_1}{p_2}$.

An encoded symbol of the applied multi layer LT code has two components; a base layer component and a refinement layer component. Thus, the degree of a symbol can be viewed as two dimensional. We denote it as $(i_B,i_R)$, where $i=i_B+i_R$. Lemma \ref{twordd} describes how to find the two-layer reduced degree distribution. The general case of $N$ layers is treated in the Appendix.

\begin{lem}(Two-Layer Reduced Degree Distribution)\label{twordd}
Given that the encoder applies $\pi(i)$ in a two-layer LT code with parameters, $\alpha$ and $\beta$, the reduced degree distribution is found as

\begin{align}
\scriptstyle \pi'(i_B',i_R') \hspace{0.1cm} = \hspace{0.1cm} & \sum_{\scriptscriptstyle i=i_B'+i_R'}^{\scriptscriptstyle i_B'+i_R'+k-L_B-L_R} \bigg(\scriptstyle\pi(i) \displaystyle\sum_{\scriptscriptstyle j=i_B'}^{\scriptscriptstyle i-i_R'} \bigg( \scriptstyle \Phi(j,i,\alpha k,k,\beta) \notag \\ 
& \scriptstyle \frac{\binom{L_B}{i_B'}\binom{\alpha k-L_B}{j-i_B'}}{\binom{\alpha k}{j}} \frac{\binom{L_R}{i_R'}\binom{(1-\alpha) k-L_R}{i-j-i_R'}}{\binom{(1-\alpha) k}{i-j}} \bigg) \bigg) \hspace{0.2cm} \mathrm{for} \hspace{0.2cm} 
\begin{smallmatrix} 0 &\le i_B' \le& L_B \\
0 &\le i_R' \le& L_R,
\end{smallmatrix} \notag \\
\scriptstyle \pi'(i_B',i_R') \hspace{0.1cm} = \hspace{0.1cm} & \scriptstyle 0 \hspace{0.2cm}\mathrm{for} \hspace{0.2cm} L_B < i_B' \le \alpha k \hspace{0.1cm} \vee L_R < i_R' \le (1-\alpha)k, \notag
\end{align}

\noindent where $\Phi$ is Wallenius' noncentral hypergeometric distribution.
\end{lem}

\begin{IEEEproof}
Consider a set of items of total size $k$, of which $\alpha k$ belong to a group of a certain property. Each individual item in this group has a factor $\beta$ higher probability of being sampled than the remaining $(1-\alpha) k$ individual items. The number of samples belonging to the group with a certain property, out of a total sample set of size $i$, is described by Wallenius' noncentral hypergeometric distribution. In this case the property is to be a part of the base layer. For the derivation of the reduced degree, it is needed to further distinguish between decoded and undecoded symbols, in both the base layer and the refinement layer. This can be modeled with regular central hypergeometric distributions, as in Lemma \ref{rdd}, since decoded and undecoded symbols are chosen with equal probabilities in the encoding process.
\end{IEEEproof}

\begin{figure}[t]
\centering
\includegraphics[width=0.95\columnwidth]{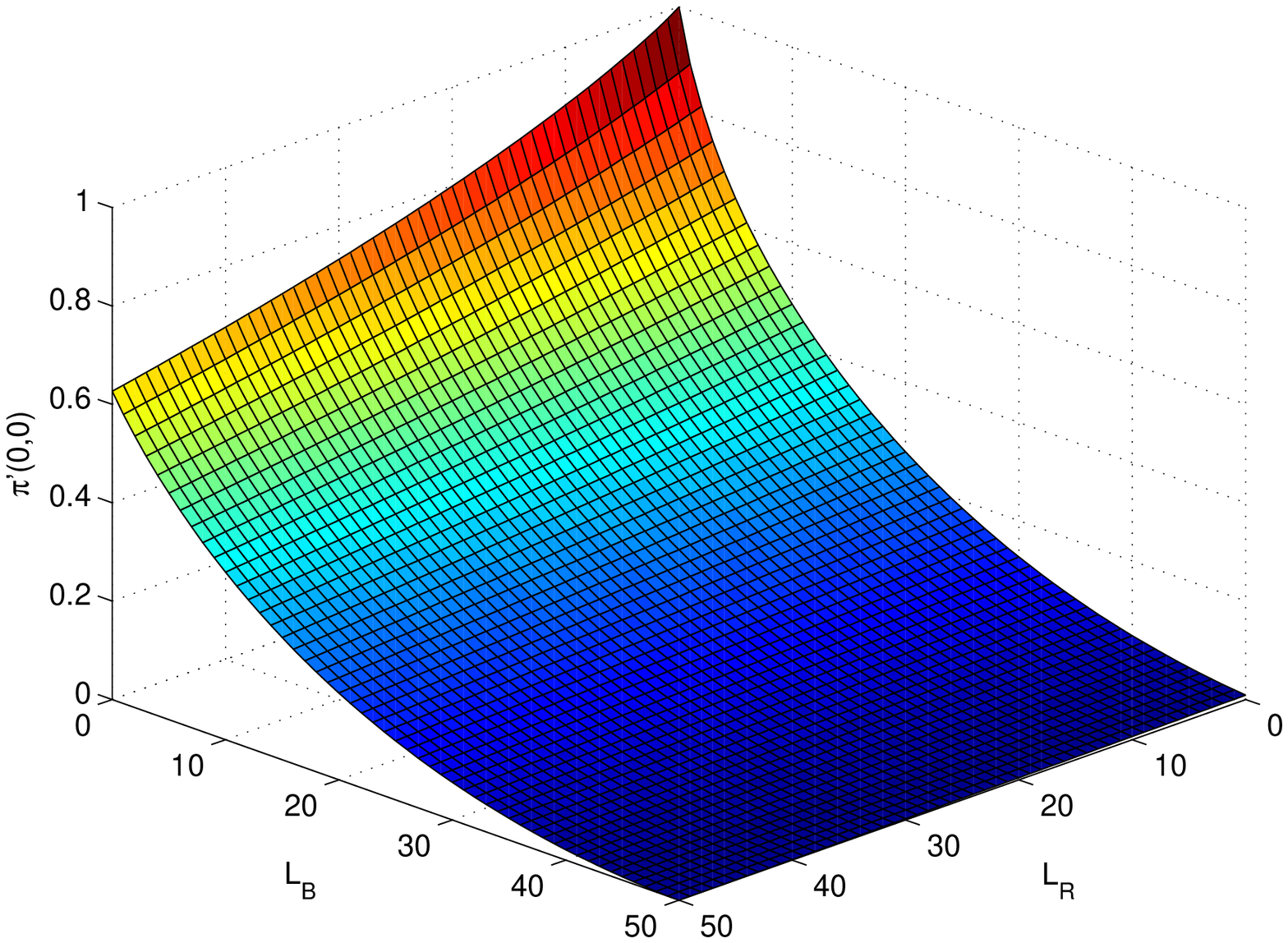}
\caption{The probability of receiving a symbol with reduced degree zero, $\pi'(0,0)$, for a two-layer LT code with parameters $\alpha=0.5$, $\beta=9$ and $k=100$.}
\label{fig:twolayerred}
\vspace{0.2cm}
\begin{minipage}[b]{0.48\columnwidth}
\centering
\includegraphics[width=\columnwidth]{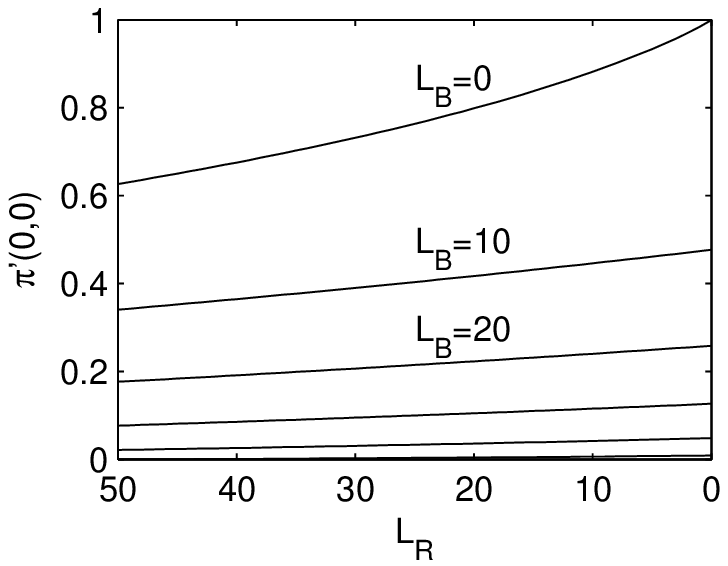}
\label{fig:proj1}
\end{minipage}
\hspace{0.0cm}
\begin{minipage}[b]{0.48\columnwidth}
\centering
\includegraphics[width=\columnwidth]{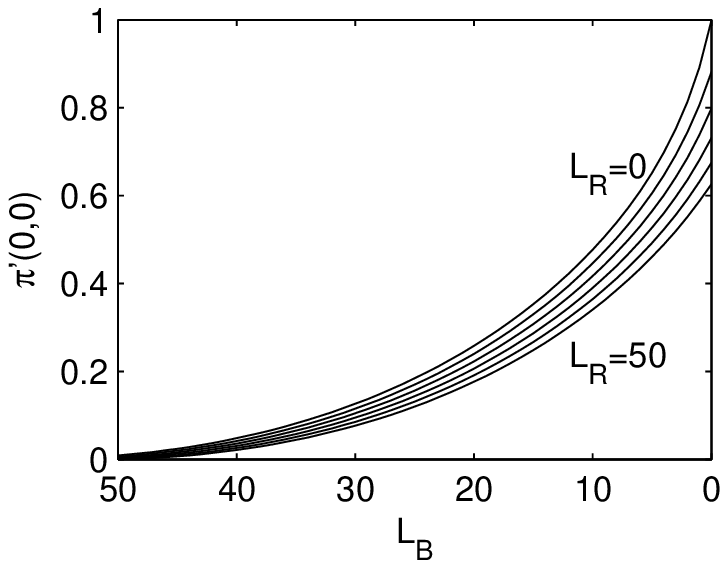}
\label{fig:proj2}
\end{minipage}
\vspace{-0.7cm}
\caption{Projections of $\pi'(0,0)$ at $L_B=\{0,10,20,30,40,50\}$ and $L_R=\{0,10,20,30,40,50\}$.}
\vspace{-0.4cm}
\end{figure}

In Fig. \ref{fig:twolayerred}, $\pi'(0,0)$ has been plotted as a function of $L_B$ and $L_R$, the number of undecoded symbols from the base layer and the refinement layer, respectively. The parameters, $\alpha=0.5$, $\beta=9$ and $k=100$ have been chosen. An optimized degree distribution is not provided in \cite{uepfc1}, thus we have chosen the RSD. The plot shows that the probability of redundancy increases faster for decreasing $L_B$ than for decreasing $L_R$, which was expected since the base layer symbols are more likely to occur as neighbors. In \cite{uepfc1} it was shown that the applied multi layer LT code provides the unequal recovery time property, which essentially means that $L_B$ decreases to zero earlier than $L_R$ during decoding. In the avalanche analogy, this means that we have two avalanches instead of one. With respect to Fig. \ref{fig:twolayerred}, it means that the probability of redundancy experienced during decoding is found along a path which starts at $L_B=L_R=50$, then moves to a point with $L_B=0$ and $L_R$ still close to $50$, and finally follows the edge of the surface towards $L_B=L_R=0$. Note the very high probability of redundancy during the entire second avalanche. This is a significant drawback of the multi layer LT code, i.e. that the decoding of refinement layers is very inefficient.

We propose to eliminate this drawback through ACKs. From the analysis of single layer LT codes, we learned that ACKs during an avalanche is not very helpful. However, in this case decoding evolves as two avalanches separated in time. This creates an opportunity to apply ACKs efficiently. The strategy is to ACK the base layer in its entirety, when it has been decoded, such that only refinement layer symbols are considered in future encoding. Note that this only requires a single feedback opportunity, which makes this scheme practical, as opposed to the single layer counterpart. We expect this application of ACKs to provide both a positive and a negative effect, as in single layer LT codes. Which one is dominant, is now studied through simulations.

\begin{figure}[t]
\centering
\includegraphics[width=0.95\columnwidth]{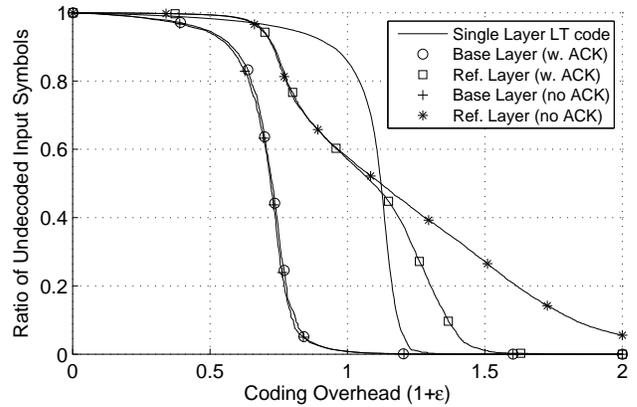}
\caption{The relative number of undecoded input symbols from the individual layers as a function of the amount of received symbols.}
\label{fig:twolayerresults}
\end{figure}

The two-layer LT code from the analysis is also chosen for the simulation, although here with $k=1000$. The code with and without ACK of the base layer are simulated. Moreover, the single layer LT code is included in order to show the effect of unequal error protection. Fig. \ref{fig:twolayerresults} shows the results. It is clear from the figure that the decoding of the refinement layer is inefficient when ACK is not applied. This is indicated by the break at the point where the base layer is decoded. However, when ACK is applied, this break is not nearly as significant, and a noticeable decrease of the necessary overhead is achieved. It can thus be concluded, that the positive effect from ACK'ing in the applied two-layer LT code outweighs the negative effect. This is in contrast to the results for the single layer LT code. In this case, we have not tried to eliminate the negative effect, as we did in the single layer case. However, it is our belief, that further potential improvements lie in the implementation of an adaptive degree distribution. This is subject to future investigations. 

\section{Numerical Results} \label{sec:results}
A simulation has been implemented in Matlab with the purpose of evaluating the investigated schemes with respect to a distortion measure. We do not simulate the distortion using actual data. Instead we apply the theoretical lower bound from rate-distortion theory \cite{ratedist}, which says that for an i.i.d., zero-mean, unit-variance Gaussian source the distortion-rate function is:

\begin{align}
d(r) = 2^{-2 r},
\label{dr}
\end{align}

\noindent where $r$ is the rate measured in bits pr. sample. We wish to evaluate the two-layer LT code from the analysis, thus we assume a lossy source code with achievable rates $r_z$, $z=0,1,2$, where $z$ denotes the number of decoded layers. The values of $r_z$ are based on the transmission of a 1 Mbit/s video stream, where the resolution is 480 times 320 at 30 frames/s. The transmitter encodes and transmits one second of video at a time, which means $r_0=0$, $r_1=\frac{\alpha 10^6}{480\cdot320\cdot30}$ and $r_2=\frac{10^6}{480\cdot320\cdot30}$. The decoding deadline of each segment of 30 frames is assumed to be the equivalent of the transmission of $2k$ symbols. The RSD is applied in all schemes, and the parameters are $k=100$, $c=0.1$, $\delta=1$, $\alpha=0.5$ and $\beta=9$. The transmission channel is modeled as a packet erasure channel, where erasures occur at symbol level at a rate denoted by $SER$. 

\begin{figure}[t]
\centering
\includegraphics[width=0.95\columnwidth]{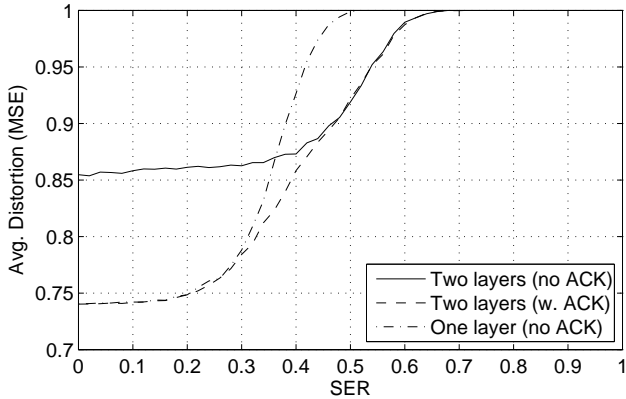}
\caption{Average distortion as a function of the symbol error rate for the investigated schemes.}
\vspace{-0.4cm}
\label{fig:rdsimul}
\end{figure}

Fig. \ref{fig:rdsimul} shows a plot of the distortion averaged over 100 seconds of ``video'' as a function of $SER$. From the plot we see that at low $SER$ and high $SER$ the two-layer LT code with ACK and the single layer LT code performs similarly. However, at $SER$ between roughly 0.3 and 0.6, the two-layer code has a significantly better performance than the single layer code. This tells us that in this region the decoding deadline is likely to occur after the early avalanche of the base layer, but before the single avalanche of the single layer code. The difference between applying ACK and not is evident at low $SER$. The high probability of redundancy makes it unlikely that the refinement layer is ever decoded. The use of ACK clearly remedies this problem. It should be noted that the performance gains shown in this simulation is achieved without any optimizations. Further potential improvements lie in the optimization of $\alpha$ and $\beta$, as well as in the implementation of an adaptive degree distribution, as suggested in section \ref{sec:analysis}.

\section{Conclusions}\label{sec:conclusion}
In this paper we have analyzed the impact of feedback in the form of acknowledgments in LT codes. Initially, standard single layer LT codes were studied. In this study, we derived the reduced degree distribution, which was shown to be a function of the number of undecoded symbols, $L$. This dependency on $L$ provides an inherent adaptive mechanism in standard LT codes. This mechanism makes LT codes able to adapt to the current decoder state, without using feedback, and is thus of crucial importance to these codes. Further analysis showed that the proposed use of acknowledgments can eliminate the probability of receiving redundant symbols. However, acknowledgments were also shown to eliminate the desirable inherent adaptive mechanism in LT codes. A modification of the degree distribution applied at the encoder was proposed, which preserves the adaptive mechanism and at the same time eliminates the probability of receiving redundant symbols. Simulations indicate a small performance improvement from the proposed code compared to standard LT codes.

We then turned our attention to multi layer LT codes, also called LT codes with unequal error protection. These were shown to be a far better match with acknowledgments. Analysis showed that the probability of receiving redundant symbols is a much bigger problem in these codes, mainly during decoding of lower prioritized layers. A simple scheme was proposed, where entire layers are acknowledged at a time. In this way, using only a single feedback message, it was possible to significantly improve the performance of a two layer LT code. Simulations also showed significant improvements compared to the standard single layer LT code.
\appendix[$N$-Layer Reduced Degree Distribution]
Lemma \ref{twordd} describes the reduced degree distribution for a two-layer LT code with unequal error protection. This distribution can be generalized to $N$ layers if the multivariate version of Wallenius' noncentral hypergeometric distribution is used instead of the univariate version from Lemma \ref{twordd}. The multivariate version describes the sampling of a set of items with $N$ different properties instead of only two. This increases the number of degrees of freedom in the distribution to $N-1$, and as a result parameters are defined as vectors. The $N$-dimensional reduced degree is denoted $\pmb{i'}$, where $i'_n$ denotes the number of undecoded neighbors belonging to the $n$'th layer. Similarly, $\pmb{L}$, $\pmb{j}$, $\pmb{\alpha}$ and $\pmb{\beta}$ are the $N$-dimensional counterparts of the parameters used in Lemma \ref{twordd}. Moreover, $\hat{i}'=\sum_{n=1}^N i'_n$, and similarly we define $\hat{L}$ and $\hat{j}$. By $\mathcal{J}$, we denote all $\pmb{j}$ which satisfy $j_n>i'n$, $n=1,2,...,N$, and $\hat{j}=i$. The N-layer reduced degree distribution can now be expressed as follows.

\begin{lem}(N-Layer Reduced Degree Distribution)\label{nrdd}
Given that the encoder applies $\pi(i)$ in an N-layer LT code with parameters, $\pmb{\alpha}$ and $\pmb{\beta}$, the reduced degree distribution is found as
\begin{align}
\scriptstyle \pi'(\pmb{i'}) \hspace{0.1cm} = \hspace{0.1cm} & \sum_{\scriptscriptstyle i=\hat{i}'}^{\scriptscriptstyle \hat{i}'+k-\hat{L}} \bigg(\scriptstyle\pi(i) \displaystyle\sum_{\scriptscriptstyle \pmb{j}\in\mathcal{J}} \bigg( \scriptstyle \Phi(\pmb{j},i,\pmb{\alpha} k,\pmb{\beta}) \displaystyle\prod_{n=1}^N \scriptstyle\frac{\binom{L_n}{i_n'}\binom{\alpha_n k-L_n}{j_n-i_n'}}{\binom{\alpha_n k}{j_n}} \bigg) \bigg)\notag \\ 
& \scriptstyle \hspace{0.2cm}\mathrm{for} \hspace{0.2cm} L_n < i_n' \le \alpha_n k, \hspace{0.2cm} n=1,2,...,N, \notag \\
\scriptstyle \pi'(\pmb{i'}) \hspace{0.1cm} = \hspace{0.1cm} & \scriptstyle 0 \hspace{0.2cm} \mathrm{elsewhere}. \notag
\end{align}
\end{lem}

\begin{IEEEproof}
See proof of Lemma \ref{twordd}.
\end{IEEEproof}



\bibliographystyle{ieeetr}
\bibliography{../../../bibliography}

\end{document}